# An Improving Method for Loop Unrolling

Meisam Booshehri, Abbas Malekpour, Peter Luksch
Chair of Distributed High Performance Computing,
Institute of Computer Science, University of Rostock,
Rostock, Germany
m_booshehri@sco.iaun.ac.ir, abbas.malekpour@uni-rostock.de, peter.luksch@uni-rostock.de

*Abstract*—In this paper we review main ideas mentioned in several other papers which talk about optimization techniques used by compilers. Here we focus on loop unrolling technique and its effect on power consumption, energy usage and also its impact on program speed up by achieving ILP (Instruction-level parallelism). Concentrating on superscalar processors, we discuss the idea of generalized loop unrolling presented by J.C. Hang and T. Leng and then we present a new method to traverse a linked list to get a better result of loop unrolling in that case. After that we mention the results of some experiments carried out on a Pentium 4 processor (as an instance of super scalar architecture). Furthermore, the results of some other experiments on supercomputer (the Alliat FX/2800 System) containing superscalar node processors would be mentioned. These experiments show that loop unrolling has a slight measurable effect on energy usage as well as power consumption. But it could be an effective way for program speed up.

## I. INTRODUCTION

Nowadays processors have the power to execute more than one instruction per clock. And this can be seen in superscalar processors. As the amount of parallel hardware within superscalar processors grows, we have to make use of some methods which effectively utilize the parallel hardware. Performance improvements can be achieved by exploiting parallelism at instruction level. Instruction level parallelism (ILP) refers to executing low level machine instructions, such as memory loads and stores, integer adds and floating point multiplies, in parallel. The amount of ILP available to superscalar processors can be limited with conventional compiler optimization techniques, which are designed for scalar processors. One of optimization techniques that in this paper we focus on it is loop unrolling which is a method for program exploiting ILP for machines with multiple functional units. It also has other benefits that we present them in section 3.

This paper is organized as follows. Section 2 describes some goals of designing a superscalar processor and the problems which would occur. Section 3 describes methods of loop unrolling and put forwards some new ideas. Section 4 reports the results of some experiments. Section 5 describes future work. Section 6 concludes. Section 7 thanks people who encouraged me to prepare this paper.

## II. SUPERSCALAR PROCESSORS

The aim of designing superscalar processors is to reduce the average of execution time per instruction through executing the instructions in parallel. To do this instruction latency should be reduced. One of cases that in designing superscalar processors we should consider it is data dependency which its side effects must be removed or at least should be minimized. This means superscalar processors must organize the results to have the computation continued correctly [2, 4].

Writing a program can be divided into several steps including writing the program code with a high-level language, translating the program to assembly code and binary code and etc. it is important to attempt to divide the program translated to assembly code, into Basic Blocks [4]. A basic block has the maximum number of instructions with a specified input and output point. Therefore, each basic block has the maximum number of successive instructions with no branch (with the exception of last instruction) and no jump (with the exception of first instruction). The basic block would always be traversed. In this manner the processor can execute a basic block in parallel. So the compilers and superscalar architecture concentrate on size of basic blocks. Through integrating some basic blocks for instance by executing Branch statements entirely, the amount of parallelism would increase. If no exception occurs within the execution time, the processor must correct all results and pipeline contents. Therefore there is a strong relation between superscalar architecture and compiler construction (especially code generator and optimizer). Certainly there are some data dependencies inside a basic block. These dependencies exist among data of various instructions. Despite RISC processors in which there are only read after write hazards, the superscalar processors may encounter read after write hazards as well as write after write hazards Because of executing instructions in parallel.

## III. GENERALIZED LOOP UNROLLING: LIMITATION AND PROPOSED SOLUTION

Loop unrolling is one kind of code transformations techniques used by compilers to reach ILP. With loop unrolling technique we transform an M-iteration loop into a loop with M/N iterations. So it is said that the loop has been unrolled N times.

**-Unrolling FOR Loops**. Consider the following countable loop:

>  for(i=0;i<100;i++)
>
>  a[i]*=2;

This FOR loop can be transformed into the following equivalent loop consisting of multiple copies of the original loop body:

>  for(i=0;i<100;i+=4){
>
>  a[i]*=2;
>
>  a[i+1]*=2;
>
>  a[i+2]*=2;
>
>  a[i+3]*=2;
>
>  }

Unlike FOR loops operating on arrays which can be unrolled simply by modifying the counter and termination condition of loop as illustrated above, WHILE loops are generally more difficult to unroll. It is so important because of difficulty in determining the termination condition of an unrolled WHILE loop. Hang and Leng et al. [1] present a method that we review it briefly.

**-Unrolling WHILE Loops.** We assume that loops are written in the form: "while **B** do **S**" the semantic of which is defined as usual. B is loop predicate and S is loop body. It is proved that the following equivalence relation holds.

$$\text{while } B \text{ do } S; \Leftrightarrow \text{while } B \text{ and } wp(S,B) \text{ begin } S; S \text{ end};$$
$$\text{while } B \text{ do } S$$

Where $\Leftrightarrow$ stands for the equivalence relation, and wp(S, B) the weakest precondition of S with respect to post condition B [3].

Therefore we can speed up the execution of the loop construct mentioned above by following steps:

1. Form wp(S,B), the weakest precondition of S with respect to B

2. Unroll the loop once by replacing it with a sequence of two loops:

    while (B and wp(S,B)) do begin S;S end;

    while B do S;

3. Simplify the predicate (B AND wp(S,B)) and the loop body **S;S** to speed up.

To illustrate, consider the following example.

**Example 1**: This example contains a loop for computing the quotient, q, of dividing b into a:

1. q=0;
2. while(a>=b)
3. {
4. a=a-b;
5. q++;
6. }

$$\Leftrightarrow$$

1. q=0;
2. While(a>=b && a>=2*b) //unrolled loop
3. {
4. a=a-b;
5. q++;
6. a=a-b;
7. q++;
8. } //end of unrolled loop
9. while(a>=b)
10. {
11. a=a-b;
12. q++;
13. }

As mentioned in [3] "The experimental results show that this unrolled loop is able to achieve a speed up factor very close to 2, and if we unroll the loop k times, we can achieve a speed up factor of k."

**Example 2:** A loop for traversing a linked list and counting the nodes traversed:

1. Count =0;
2. While (lp!=NULL)
3. {
4. lp=lp->next;
5. Count++;
6. }

The best solution presented by Hang and Leng [3] is to attach a special node named NULL_NODE at the end of the list. The link field of this node points to the node itself.

With this idea, after unrolling the loop twice, it becomes:

1. Count=0;
2. lp1=lp->next;
3. lp2=lp1->next;

4. While(lp2!=NULL)

5. {

6. Count+=3;

7. lp=lp2->next;

8. lp1=lp->next;

9. lp2=lp1->next;

10. }

11. While(lp!=NULL)

12. {

13. lp=lp->next;

14. Count++;

15. }

The instructions number 6,7,8,9 forms a basic block, but because of data dependencies superscalar processors can not execute these instructions in parallel. The benefits of this unrolled loop come from less loop-overhead and not from ILP. So we suggest a new way to solve this problem (that is traversing linked list and counting its nodes). And we hope the new method could increase level of parallelism. This is not a general solution and just solves this problem; however, this gives us a new idea of increasing pointers to traverse the list from different positions. The solution is as follows.

**Proposed Solution:** We use a two-way linked list which also has two pointers named **first** (pointing to the first node) and **last** (pointing to the last node). So we have the following algorithm:

1. F=first;

2. L=last;

3. Count=0;

4. While ((F!=L) || (F->right!=L))

5. {

6. F=F->right;

7. L=L->left;

8. Count+=2;

9. }

10. If(F=L)

11. Count-=1;

In this algorithm we encounter two possible states as comes below:

1. **The number of list nodes is odd.** In this state when the pointers F and L move to the middle of list, they finally visit the middle node of list at the same time. Therefore the termination condition of loop is F=L and the middle node won't be counted. So we count the node by using the last two instructions.

2. **The number of list nodes is even.** In this state the pointers F and L finally reach the state in which following relations holds:

(F->right==L) and (L->left==F)

So one of these conditions could be used to form the termination condition.

IV. POWER CONSUMPTION, ENERGY USAGE AND SPEED UP

-**Simulation or measuring.** The program code plays an effective role in power consumption of a processor. So some research has been done studying the impact of compiler optimizations on power consumption. Given a particular architecture the programs that are run on it will have a significant influence on the energy usage of the processor. The relative effect of program behavior on processor energy and power consumption can be demonstrated in simulation. But there are some factors such as clock generation and distribution, energy leakage, power leakage and etc. that make it difficult to have an accurate architecture-level simulation to give us enough information about the effect of a program on a real processor [1]. Therefore, we have to measure the effect of a program on a real processor and not just in simulation.

-**Results.** Here we review the results of some experiments done to study impact of loop unrolling technique on three factors: power consumption and energy usage of a superscalar processor, and also program speed up. Seng and Tullsen et al.[1] study the effect of loop unrolling on power consumption and energy usage. They measure the energy usage and power consumption of a 2.0 GHZ Intel Pentium 4 processor. They run different benchmarks compiled with various optimizations using the Intel C++ compiler and quantify the energy and power differences when running different binaries. They conclude that "when applying loop unrolling, there is a slight measurable reduction in energy, for little or no effect on performance. For the binaries where loop unrolling is enabled, the total energy is reduced as well as the power consumption. The difference in terms of energy and power is very small, though."

Mahlke et al. [2] study the effect of loop unrolling as a technique to reach ILP on supercomputers which contains superscalar node processors. They reach the result that "with conventional optimization taken as a baseline, loop unrolling and register renaming yields an overall average speed up of 5.1 on an issue-8 processor". The maximum number of instructions that an issue-8 processor can fetch and issue per cycle is 8. The other result that they've reached is that the ILP transformations including loop unrolling increase the register usage of loops.

## V. CONCLUSION

In this study we review the ideas mentioned in several other papers which talk about compiler optimization techniques. Focusing on loop unrolling and superscalar architecture, we discuss the idea of generalized loop unrolling presented by J.C. Hang and T. Leng and then we present a new method to traverse a linked list to get a better result of loop unrolling in that case. After that with comparing and examining ideas we reach some results as follows. Loop unrolling has a slight measurable effect on energy usage as well as power consumption by which no huge change in performance would occur. But it could be an effective method for program speed up. An important issue is that the loop unrolling technique generally won't bring the expected performance to the programs without other optimization techniques such as register renaming. These results have been gained by using measuring technique accompanying simulation technique.

## VI. FUTURE WORK

Additional work that we would like to perform would be to change existing algorithms which works on data structures like linked list or present some new ones to reduce the probability of occurring hazards (like read after write hazards) that force the compilers to shorten the size of basic blocks and then not using the superscalar processors' ability, effectively. In other words, we want to optimize the way of writing code for data structures to reach some standard rules of programming which result in using superscalar architecture, effectively. Or we can give this task to compilers (and not programmers) to use some standard rules in code transformations. Or we may reach a tradeoff between programmers and compilers to use some standard rules. Another thing that we guess is that the rules which we want to use may conflict some software engineering considerations in programming. So another trade off also is needed here.

## I. Authors' information

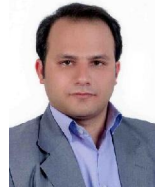

**Meisam Booshehri** was born in Iran. He received his Master Degree in Software Engineering from IAUN in 2012. Currently, he is a lecturer at Payame Noor University (PNU), Iran. He is also a member of Young Researchers Club, Sepidan Branch, Islamic Azad University, Sepidan, Iran. His research interests include parallel and distributed computing, Compilers and Semantic Web.
Email: m_booshehri@sco.iaun.ac.ir

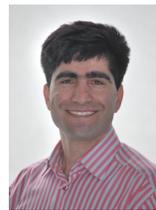

**Abbas Malekpour**[*] is currently an Assistant Professor in the Institute of Distributed High Performance Computing at University of Rostock. He received his Master Degree from Stuttgart University and his Ph.D. degree from University of Rostock, Germany. From 2002 to 2004 he was with Institute of Telematics Research Group at university of Karlsruhe, Germany. And from 2004 to 2010, he was a research assistant in MICON Electronics and Telecommunications Research Institute at University of Rostock, Germany. His current research interests include the areas of Mobile and Concurrent Multi-path Communication prototyping.

[*] Corresponding Author at: Chair of Distributed High Performance Computing, Institute of Computer Science, University of Rostock, Rostock, Germany
Email: abbas.malekpour@uni-rostock.de

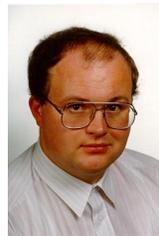

**Peter Luksch** finished his study in computer science and received his Ph.D. degree in Parallel Discrete Event Simulation on Distributed Memory Multiprocessors from Technische Universität München, Germany, in 1993. Currently, he is a Professor at University of Rostock and Head of the Chair of Distributed High Performance Computing. During the years 1993 to 2003 he was a Senior Research Assistant and Lecturer at LRR-TUM at TUM. He finished his Postdoctoral Lecture Qualification (Habilitation) in Increased Productivity in Computational Prototyping with the Help of Parallel and Distributed Computing in 2000. His current research topics include parallel and distributed computing and computational prototyping.
Email: peter.luksch@uni-rostock.de